\documentclass[12pt,thmsa]{article}
\usepackage{amsfonts}

\input{tcilatex}
\begin{document}

\author{Karl-Georg Schlesinger \\
e-mail: kggschles@gmail.com}
\title{Eight dimensional physics and the Langlands program -- A short note}
\date{11.11.2011}
\maketitle

\begin{abstract}
We argue that a special step in the chain of dualities used in \cite{Tan} implicitly suggests to view Langlands duality as being fundamentally rooted in an eight-dimensional theory on the F-theory 7-brane. We give further arguments why such an eight-dimensional perspective might be of interest.

\end{abstract}

\bigskip

As was shown in the fundamental work of \cite{KW}, geometric Langlands duality for algebraic curves would be a consequence of $S$-duality of $N=4$ SYM-theory in $d=4$ space-time dimensions if the latter could be established. An even deeper perspective on geometric Langlands duality is offered by the six dimensional view as advocated in \cite{Wit}. String theory conjectures the existence of a $d=6$ conformal field theory of type $(2,0)$ supersymmetry which can be seen e.g. as an effective limit of the worldvolume theory of the $NS5$-brane in type $IIA$ string theory. Suppose this conformal field theory to live on a space-time $X_6$ with

\begin{equation}
X_6=X_4 \times T^2
\end{equation}

In the limit

\begin{equation}
vol(T^2) \ll vol(X_4)
\end{equation}

we get $N=4$ SYM-theory on $X_4$ by dimensional reduction and the $S$-duality of this theory is a \textit{consequence} of the natural action of $SL(2,\mathbb{Z})$ on the torus $T^2$. So, while in $d=4$ we would have to prove $S$-duality for the $N=4$ SYM-theory, in order to derive geometric Langlands duality for algebraic curves, in $d=6$ it would suffice to prove the very existence of the $(2,0)$ superconformal field theory.  This means that the $d=6$ theory appears to be the geometric structure which turns $S$-duality (and hence geometric Langlands duality) into a \textit{manifest} symmetry, much the same way in which differential geometry turns general covariance into a manifest symmetry. We tacitly assume, here, and in the sequel that the gauge group $G$ of the $N=4$ SYM-theory is from the $ADE$-series since this is the most general type of $G$ which can be associated to the $d=6$ superconformal theory. \\
But the $d=6$ perspective achives even more: Consider the space-time 

\begin{equation}
X_6=\Sigma \times Y_4
\end{equation}
 
with $\Sigma$ a Riemann surface and $Y_4$ a compact Hyperk{\"a}hler manifold of (real) dimension 4. In the limit 

\begin{equation}
vol(Y_4) \ll vol(\Sigma)
\end{equation}

the $d=6$ superconformal field theory reduces to a sigma model on $\Sigma$ with the target space given by the instanton moduli space $Inst_G(Y_4)$, i.e. the space of (anti-) self-dual $G$-connections on $Y_4$. \\
As was shown in \cite{Tan} and \cite{Wit}, one can use this dimensional reduction of the $d=6$ theory to derive the results of \cite{BF} and \cite{Nak} on geometric Langlands duality for the algebraic surface $Y_4$, for $G$ from the $A$-series, and extend them to the $D$-series case. As shown in \cite{Tan}, one uses a chain of dualities of string- and $M$-theory , leading to a duality of two different $M$-theory backgrounds involving $M5$-branes (backgrounds (3.1) and (3.8) in \cite{Tan}), in order to establish the geometric Langlands duality for the algebraic surface $Y_4$.\\
But the argument of \cite{Tan} offers a far wider potential: From a string- and $M$-theory perspective, each single step in the whole chain of backgrounds ((3.1) to (3.8) in \cite{Tan}) is a duality transformation, not only the whole chain from the first to the last background. So, this chain of dualities embeds geometric Langlands duality for algebraic surfaces into a complete chain of dualities which should offer very different reformulations of the structures of the geometric Langlands program. E. g. background (3.7) is a background in type $IIB$ string theory. Usually -- as we have sketched above -- the geometric Langlands program is associated to the $(2,0)$ superconformal field theory which is the effective limit of the little stringt theory (LST) on the $NS5$-brane in type $IIA$ (or the $M5$-brane in $d=11$). But $T$-duality links this theory to the LST on the $NS5$-brane in type $IIIB$ which has a gauge theory with $(1,1)$ supersymmetry as its effective limit. The chain of dualities of \cite{Tan}
means that the geometric Langlands data should also have a formulation in this setting. Especially, one step in the chain of dualities is the application of $S$-duality of type $IIB$ string theory, which leads from LST on the $NS5$-brane in type $IIB$ to LST on the $D5$-brane, i.e. a formulation of the $(1,1)$ theory on the $D5$. As was discussed in \cite{Dij}, in this setting we can reduce the $(1,1)$ theory to a sigma model and get, once again, the instanton moduli space on the four dimensional compactification space as target space. This means that the dual formulation with the $(1,1)$ theory should involve a duality on $Inst_G(Y_4)$ for the data of the geometric Langlands program. \\
Now, let us focus on the special step in the chain of dualities where one applies $S$-duality of type $IIB$ to pass from an $NS5$-brane to a $D5$-brane. This is an especially interesting step for several reasons. First, as for geometric Langlands duality of algebraic curves, one could also ask for the case of geometric Langlands duality of algebraic surfaces how to make the duality a manifest symmetry. For the step of applying $S$-duality of type $IIB$ in the chain of dualities, there is at least an idea how one should achive this. Very much as in the transition from $N=4$ SYM-theory in $d=4$ to $d=6$ $(2,0)$ superconformal field theory, one would like to make the $S$-duality group $SL(2,\mathbb{Z})$ manifest by multiplying the $d=10$ space-time of type $IIB$ with a torus $T^2$ (or, more generally, consider a $d=12$ space-time, fibered by an elliptic curve), passing from type $IIB$ string theory to a $d=12$ theory. The LST on the $D5$ or $NS5$ in type $IIB$ should in this way arise from a $d=8$ theory on the worldvolume of a 7-brane in this $d=12$ theory. This is the original idea of $F$-theory, as proposed in \cite{Vaf}. It is meanwhile clear that $F$-theory does not work in this simple form since it is an inherently non-perturbative theory (see e.g. \cite{Wei} for a review). Nevertheless, it is still the expectation that a fundamental formulation of $F$-theory would be a manifestly $S$-duality invariant formulation of type $IIB$ string theory. \\
A second reason why this $d=8$ theory on the $F$-theory 7-brane should be interesting is that it is known that from an $M$-theory perspective $S$-duality of type $IIB$ string theory is a remnant of covariance in $d=11$. How to make the $M5$ contribution of $M$-theory (i.e. the 90\% part of the theory, by counting central charges of the $d=11$ superalgebra) covariant is one of the central open questions of $M$-theory (as making the graviton contribution covariant leads one to the full Einstein equations). An $S$-duality invariant $d=8$ formulation of the $(1,1)$ LST of type $IIB$ string theory might possibly shed some light on this question. It is especially interesting that in this way the central question of covariance of $M$-theory ties in with symmetries related to Langlands duality. \\
A third point arises from the search for a manifest formulation of Langlands duality of algebraic curves, as discussed above. Though the $(2,0)$ superconformal field theory in $d=6$ arises as an effective limit, it is non-renormalizable. From a field theory perspective the belief is (see e.g. \cite{Aha} for a general review on the subject of LSTs and their limits) that, though the $d=6$ theory does not have a UV fixed point as a quantum field theory, the $d=6$ LST on the $NS5$-brane in type $IIA$ string theory should provide the UV fixed point (in the sense of a UV completion of the theory). But this means that the effective superconformal field theory can probably not be the basis of a mathematically rigorous formulation of a manifest notion of Langlands duality of algebraic curves (since it is not expected to be a consistent theory) but a mathematically rigorous formulation should ultimately require the full LST (We will in this note ignore the point that ultimately the LST -- if we take the $M5$-brane perspective on it -- might require embedding into full $M$-theory since the partition function -- because of the self-duality condition on the 2-form on the $M5$-brane -- seems to require coupling to the $d=11$ $C$-field, in order to make it well defined, see \cite{Wit 1996}). Since $S$-duality of type $IIB$ string theory is a strong-weak coupling duality, a mainfest $d=8$ formulation might, again, be appropriate to shed some light on the UV-completion question. So, even for a mathematically rigorous \textit{and} manifest formulation of Langlands duality of algebraic curves, a $d=8$ theory on the 7-brane in $F$-theory should be an interesting starting point. \\
Let us remark at this point that it is at least known that the special topological twist of $N=4$ SYM-theory in $d=4$, needed for geometric Langlands duality of algebraic curves, can be gained from a $d=8$ theory, with an interesting relation to nonabelian Seiberg-Witten theory arising in this way (see \cite{BKS}). \\

Let us close this note by taking a very brief look on some additional topics which might offer further support for such a $d=8$ $F$-theory 7-brane perspective: 

\begin{itemize}

\item The gravitational instantons of Taub-NUT geometry, which are so decisive in the chain of dualities of \cite{Tan} (and gauge instantons on these gravitational instantons, as they appear with the moduli spaces $Inst_G(Y_4)$), also arise in a completely different context in the setting of $F$-theory 7-branes (see \cite{GS}). The work of \cite{GS} hints at a relation of these structures to transdimensional tunneling and possibly an inflationary scenario within $F$-theory. Also, the $d=8$ worldvolume theory on the $F$-theory 7-brane strongly relates to string theory phenomenology, embedding the standard model into string theory. Again, in this work structures well known from the geomtetric Langlands side, like Higgs bundles and Hitchin's equations play a deep role (see \cite{DW}, \cite{PW}). 

\item There is not only a deep mathematical relation between the landscape of string theory vacua, as it emerges from $F$-theory, and the (string theoretic) microscopic description of black holes but this is rooted in the large entropy of black holes (see \cite{Den}, \cite{DM}). On the other hand, black hole attractors show the appearance of arithmetic varieties over number fields (see \cite{Moo}). Hitchin moduli space does not only take a center stage role in the gauge theory approach to the geometric Langlands program, initiated in \cite{KW}, but also in the classical (arithmetic) Langlands program (in the recent proof of the fundamental lemma, see \cite{Ngo 2006a}, \cite{Ngo 2006b}). Since the classical Langlands program uses varieties over number fields, such similarities have often remained just that but the black hole attractor mechanism might be a hint that there could be a common structure, rooted in physics, behind the two versions of the Langlands program (as was speculated in \cite{Ati}). 

\item Finally, recent work (see \cite{Tan 2011}) has shown that for a detailed mathematical picture of geometric Langlands duality it is necessary to include supersymmetry breaking effects. On the other hand, breaking of supersymmetry is an important topic in the study of the $F$-theory landscape.  

\end{itemize}

Of course, our arguments are far from conclusive to demonstrate that a $d=8$ approach to the Langlands program would be fruitful. But an $F$-theory 7-brane perspective seems to be at the center stage of a whole bunch of similar structures, appearing in very different areas (from classical and geometric Langlands to inflation and the standard model). In a talk on a -- very different -- matter of coincidences (and the question if one should search for a deeper explanation of them) Dennis Sciama asked the auditorium: Would you raise your eyebrows? Would you?


\bigskip

\begin{thebibliography}{xxx}

\bibitem[Aha]{Aha} O. Aharony, \textit{A brief review of little string theories}, hep-th/9911147.

\bibitem[Ati]{Ati} M. Atiyah, \textit{Edinburgh lectures on geometry, analysis and physics}, arXiv:1009.4827.

\bibitem[BF]{BF}  A. Braverman, M. Finkelberg, \textit{Pursuing the double
affine Grassmannian I: Transversal slices via instantons on }$A_{k\text{ }}$%
\textit{singularities}, arXiv:0711.2083.

\bibitem[BKS]{BKS} L. Baulieu, H. Kanno, I. M. Singer, \textit{Special quantum field theories in eight and other dimensions}, hep-th/9704167v2.

\bibitem[Den]{Den} F. Denef, \textit{Les Houches lectures on constructing string vacua}, arXiv:0803:1194.

\bibitem[Dij]{Dij}  R. Dijkgraaf, \textit{Instanton strings and
Hyperk\"{a}hler geometry}, hep-th/9810210.

\bibitem[DM]{DM} F. Denef, G. W. Moore, \textit{Split states, entropy enigmas, holes and halos}, hep-th/0702146.

\bibitem[DW]{DW} R. Donagi, M. Wijnholt, \textit{Higgs bundles and UV completion in F-theory}, arXiv:0904.1218.

\bibitem[GS]{GS} T. W. Grimm, R. Savelli, \textit{Gravitational instantons and fluxes from M/F-theory on Calabi-Yau fourfolds}, arXiv:1109.3191.

\bibitem[KW]{KW}  A. Kapustin, E. Witten, \textit{Electric-magnetic duality
and the geometric Langlands program}, hep-th/0604151.

\bibitem[Moo]{Moo} G. W. Moore, \textit{Arithmetic and attractors}, hep-th/9807087.

\bibitem[Nak]{Nak}  H. Nakajima, \textit{Quiver varieties and branching},
arXiv:0809.2605.

\bibitem[Ngo 2006a]{Ngo 2006a}  B.-C. Ng$\widehat{o}$, \textit{Fibration de
Hitchin et endoscopie}, Invent. Math. \textbf{164} (2006), 399-453.

\bibitem[Ngo 2006b]{Ngo 2006b}  B.-C. Ng$\widehat{o}$, \textit{Le lemme
fondamental pour les algebres de Lie}, preprint, available at
http://www.math.u-psud.fr/\symbol{126}ngo/LFLS.pdf

\bibitem[PW]{PW} T. Pantev, M. Wijnholt, \textit{Hitchin's equations and M-theory phenomenology}, arXiv:0905.1968.

\bibitem[Tan 2008]{Tan}  M.-C. Tan, \textit{Five-branes in M-theory and a
two-dimensional geometric Langlands duality}, arXiv:0807.1107.

\bibitem[Tan 2011]{Tan 2011} M.-C. Tan, \textit{Quasi-topological gauged sigma models, the geometric Langlands program, and knots}, arXiv:1111.0691.

\bibitem[Vaf]{Vaf} C. Vafa, \textit{Evidence for F-theory}, hep-th/9602022.

\bibitem[Wei]{Wei} T. Weigand, \textit{Lectures on F-theory compactifications and model building}, arXiv:1009:3497.

\bibitem[Wit 1996]{Wit 1996}, E. Witten, \textit{Five-brane effective action in M-theory}, hep-th/9610234. 

\bibitem[Wit 2009]{Wit}  E. Witten, \textit{Geometric Langlands from six
dimensions}, arXiv:0905.2720.

\end{thebibliography}
\end{document}